\documentclass[aps,prb,floatfix,amsmath,amssymb,eqsecnum,nofootinbib,twocolumn]{revtex4}
\usepackage{graphicx}
\usepackage{dcolumn}
\usepackage{bbm}
\usepackage{epstopdf}
\usepackage[bookmarks=true,colorlinks,linkcolor=blue,urlcolor=blue,citecolor=blue]{hyperref}
\usepackage{color}

\usepackage{setspace} 
\newcommand{\beq}{\begin{equation}}
\newcommand{\eeq}{\end{equation}}
\newcommand{\beqn}{\begin{eqnarray}}
\newcommand{\eeqn}{\end{eqnarray}}

\usepackage{amsmath}
\usepackage{amssymb}

\def \q{{\mathbf{q}}}

\def \Q{{\mathbf{Q}}}
\def \k{{\mathbf{k}}}

\def \q{{\mathbf{q}}}
\def \p{{\mathbf{p}}}

\begin{document}

\title{Nematic fluctuations and their wave vector in two-dimensional metals}

\author{Matthias Punk}
\affiliation{Institute for Quantum Optics and Quantum Information, Austrian Academy of Sciences, 6020 Innsbruck, Austria}
\affiliation{Institute for Theoretical Physics, University of Innsbruck, 6020 Innsbruck, Austria }

\date{\today}

\begin{abstract}We revisit the problem of electrons on a square lattice below half filling close to an Ising-nematic quantum critical point. For Fermi surfaces with sufficiently strong antinodal nesting, the static nematic susceptibility is maximal at the antinodal nesting wave vector within a simple RPA calculation. We present a detailed analysis of the nematic susceptibility within Eliashberg theory and show that the strong interaction between Fermions in the antinodal regions shifts the maximum of the nematic susceptibility to slightly larger wave vectors. The corresponding order is akin to the incommensurate charge density wave with d-wave form factor found recently in some underdoped cuprate materials.
At sufficiently high temperatures around $T/t \sim 0.1$ nematic fluctuations are strongest at zero wave vector.
\end{abstract}

\maketitle

\section{Introduction}

Incommensurate charge density wave (CDW) order has been observed recently in several cuprate high-T$_c$ superconductors at small hole doping.\cite{Wu2011, Harrison2011, Ghiringhelli2012, Chang2012, Achkar2012, LeBoeuf2013, daSilva2014} The measured ordering wave vectors $\Q=(\pm Q_0,0)$ and $(0,\pm Q_0)$ are oriented along the principal axes of the CuO$_2$ planes and take values around $Q_0 \sim 0.25 \dots 0.32$ in reciprocal lattice units, corresponding to a modulation of the CDW order parameter with a period between three and four lattice spacings. Three recent experiments by Comin \emph{et al.}~[\onlinecite{Comin}], Fujita \emph{et al.}~[\onlinecite{Fujita}] and Achkar \emph{et al.}~[\onlinecite{Achkar2014}]  showed that this CDW order has a predominant d-wave form factor in several cuprate families such as YBa$_2$Cu$_3$O$_{6+y}$,  Bi$_2$Sr$_2$CaCu$_2$O$_{8+x}$ and Ca$_{2-x}$Na$_x$CuO$_2$Cl$_2$. 

This observation of CDW order has prompted theoretical interest in the origin of the axial ordering wave vectors $\Q=(\pm Q_0,0)$ and $(0,\pm Q_0)$. Interestingly, many theoretical approaches predict a dominant instability towards a CDW with d-wave form factor at diagonal wave vectors $(\pm Q_0,\pm Q_0)$, whereas the experimentally observed axial wave vectors appear only as a subleading instability.\cite{Metlitski, Holder, Sachdev, Efetov, Bulut, Sau, Carvalho, Thomson} Other theoretical works argue that the dominant instability is indeed at the axial ordering wave vector, either because of the interplay between CDW and superconductivity,\cite{Wang, Debanjan} the effect of a sufficiently large nearest-neighbor Coulomb repulsion $V$ in a $t$-$J$-$V$ model,\cite{Allais} or because of other microscopic details.\cite{Davis, Melikyan,Atkinson,Yamakawa}

The observed charge-density wave with a d-wave form factor is akin to an incommensurate Ising-nematic order. For this reason we revisit the problem
of two-dimensional metals close to an Ising-nematic quantum critical point (QCP), which has been studied thoroughly in the past,\cite{Yamase, Halboth, Oganesyan, Metzner, Kee, Yamase2005, DellAnna, Rech, Lawler, Garst, Yamase2012, Drukier} with the aim to determine the wave vector at which nematic fluctuations are strongest. In fact, a similar analysis has been performed earlier within a random phase approximation (RPA) by Holder and Metzner,~\cite{Holder} where it was shown that the static nematic susceptibility is largest along $2 k_F$ lines, i.e.~at momenta connecting points on the Fermi surface with parallel tangents. More precisely, they found
that the nematic susceptibility for electron densities larger than van-Hove filling\cite{footnote} is maximal at diagonal wave vectors $\Q=(\pm Q_0,\pm Q_0)$ where two $2 k_F$ lines cross. Again, these results suggest that the ordering along axial wave vectors is a subleading instability.

Here we demonstrate that the static nematic susceptibility can also be maximal at axial wave vectors $\Q=(\pm Q_0,0)$ and $(0,\pm Q_0)$, depending on the nesting properties (i.e.~the curvature) of the Fermi surface in the antinodal region. Indeed, for sufficiently strong nesting of the Fermi surface in the antinodal region, the maximum of the nematic susceptibility is precisely at the antinodal nesting wave vector within a simple RPA calculation. 

Even though the antinodal nesting wave vector is closer to the experimentally observed ordering wave vector than the diagonal one, it misses one essential feature.
Experiments suggest that the CDW ordering wave vector connects the tips of the Fermi arcs in the pseudo-gap regime, rather than the putative Fermi surface in the antinodes.\cite{Comin2014} This observation might be suggestive of the fact that the CDW order is a secondary instability of a genuine pseudogap phase which already features a reconstructed Fermi surface.\cite{Lee, Debanjan2}

Here we point out one possible resolution how the observed ordering wave vector could be reconciled with the picture of a CDW instability of an ordinary Fermi liquid with a large Fermi surface. 
In the nematic QCP scenario electrons close to the antinodes interact strongly and are no well-defined quasiparticles. Nematic fluctuations are strongest at $2 k_F$ lines where quasiparticle coherence is important, however. It is thus conceivable that the maximum of the nematic susceptibility shifts to larger wave vectors which connect points on the Fermi surface away from the antinodes where the quasiparticle coherence is larger, even though these points have no perfectly parallel tangents. In this paper we present results from an Eliashberg-type calculation of the nematic susceptibility which suggest that this is indeed the case, even though this effect is rather small. The wave vector is only slightly shifted away from the antinodes, and we conclude that the nematic QCP scenario alone can hardly account for the experimentally observed CDW ordering wave vector. Superconducting fluctuations, which are clearly important in the context of cuprates, likely increase this shift of the CDW ordering wave vector, but a detailed study of this effect is beyond the scope of this paper.

Another interesting result from the Eliashberg computation is that the maximum of the nematic susceptibility shifts to zero wave vector at sufficiently high temperatures above $T/t \sim 0.1$. The incommensurate CDW fluctuations at low temperatures thus appear as ordinary, intra-unit-cell nematic fluctuations at higher temperatures. Indeed, intra-unit-cell nematicity has been reported in cuprates below the pseudogap temperature.\cite{Ando, Kohsaka, Hinkov, Daou, Lawler2} However, due to the gradual onset of CDW order as a function of temperature
it is not clear at present if these observations are related, or if a regime with intra-unit cell nematicity exists above the onset of CDW order.\cite{Julien}

\begin{figure}
\begin{center}
\includegraphics[width=0.65 \columnwidth]{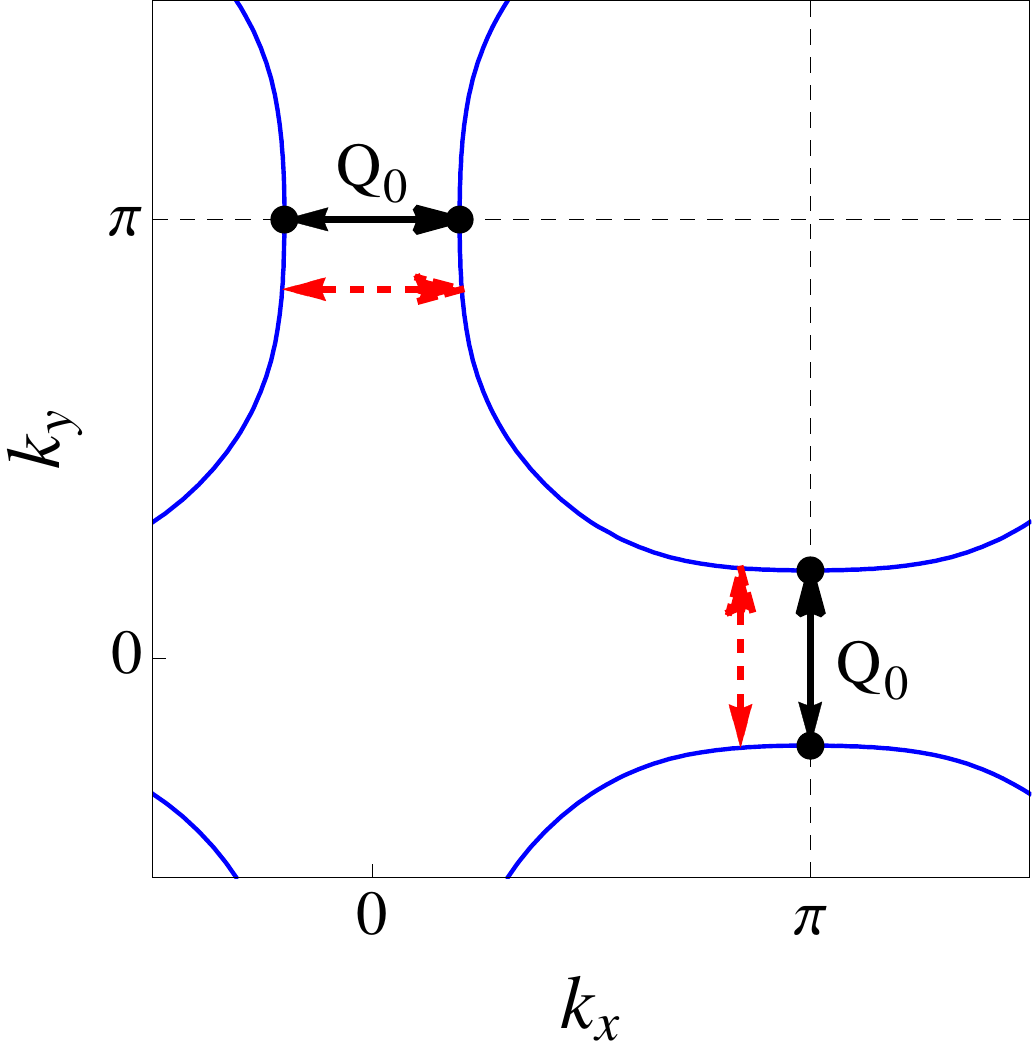}
\caption{(Color online) Ordering wave vectors and Fermi surface. The black arrows indicate the antinodal nesting wave vectors, where the static nematic susceptibility is maximal within RPA for Fermi surfaces with sufficiently strong antinodal nesting. In Eliashberg theory the maximum of the susceptibility is shifted to slightly larger wave vectors away from the antinodes, indicated by the red dashed arrows.}
\label{fig:FS}
\end{center}
\end{figure}

The rest of the paper is outlined as follows: in Sec.~\ref{sec:model} we introduce the model of electrons on a square lattice coupled to Ising-nematic order parameter fluctuations. Sec.~\ref{sec:RPA} discusses RPA results for the nematic susceptibility and in Sec.~\ref{sec:eliashberg} we present results from Eliashberg theory. 

\section{Model}
\label{sec:model}

We start from the model of electrons on a square lattice coupled to an Ising-nematic order parameter introduced in Ref.~[\onlinecite{Hartnoll}], which is defined by the euclidean action
\begin{eqnarray}
S&=&\sum_q c^*_{q \sigma} (-i \omega_n +\xi_\q) c_{q \sigma} + \sum_k \frac{s}{2} \, \phi_k \phi_{-k} \notag \\ 
&&+ \lambda \sum_{k,q} V_{\k,\q} \, \phi_{k} \, c^*_{q+k/2, \sigma} c_{q-k/2, \sigma}  \ .
\label{eq:model}
\end{eqnarray}
Here the real field $\phi_k$ denotes the Ising-nematic order parameter and $s$ defines the distance to the critical point. Electrons with Matsubara frequency $\omega_n$, momentum $\q$ and spin $\sigma$ are denoted by the fermionic fields $c_{q\sigma}$, where we use the shorthand notation $q=(\omega_n,\q)$ and $\sum_q = \beta^{-1} \sum_{\omega_n} \int_{\text{BZ}} d^2q/(2 \pi)^2$ and $\lambda$ is the interaction strength. 
In the following we use a standard tight binding dispersion for the electrons with up to third-nearest neighbor hopping on a square lattice, which takes the form
\begin{eqnarray}
\xi_\q &=& -2 t (\cos q_x +\cos q_y) - 4 t'  \cos q_x \, \cos q_y  \notag \\
&&-2 t'' (\cos 2 q_x + \cos 2 q_y) -\mu \ ,
\end{eqnarray} 
where $\mu$ is the chemical potential (the lattice constant is set to unity throughout this paper and in the following we measure energy in units of $t=1$).

On a square lattice the Ising nematic order has $d_{x^2-y^2}$ symmetry and correspondingly the form factor $V_{\k,\q}$ in the interaction term of Eq.~\eqref{eq:model} is typically chosen as $V_{\k,\q} \sim d_\q = \cos q_x - \cos q_y$. In this case the nematic order parameter at finite $\k$ is identical to a bond-density-wave order with d-wave form factor which has been discussed in the context of CDW ordering in underdoped cuprates.\cite{Sachdev, Fujita} By contrast, in the model introduced in Ref.~[\onlinecite{Hartnoll}] the Ising field $\phi_i$ lives on the sites $i$ rather than the bonds of the square lattice. In the context of hole-doped cuprates this site-centered Ising field can be viewed as a deformation of a Zhang-Rice singlet which breaks $90^\circ$ rotation invariance.\cite{ZhangRice} The form factor $V_{\k,\q}$ in Eq.~\eqref{eq:model} corresponding to such a site-centered field $\phi_i$ takes the slightly different form
\begin{eqnarray}
V_{\k,\q} &=&  d_{\q+\k/2} + d_{\q-\k/2} \notag \\
&=& 2 \left[ \cos q_x \cos \frac{k_x}{2} - \cos q_y \cos \frac{k_y}{2} \right] \ ,
\label{eq:formfactor}
\end{eqnarray}
and depends explicitly on the fluctuation wave vector $\k$, but obviously reduces to the standard definition at $\k=0$. In the following we will always use Eq.~\eqref{eq:formfactor} in our computations. Note that this form factor respects all symmetries of an Ising-nematic order parameter with $d_{x^2-y^2}$ symmetry by construction, i.e.~it is odd under $\pi/2$-rotations and reflections about the nodal directions, and it is even under time-reversal as well as reflections about the x- and y-axis.
Together with the Ising symmetry $\phi_k \to -\phi_k$ the action \eqref{eq:model} is invariant under the square lattice point group. At the level of the bare action the Ising symmetry is broken for $s<0$, in which case the point group symmetry is reduced from tetragonal to orthorhombic. 
At a fixed ordering wave vector $\k$ the form factor \eqref{eq:formfactor} can be viewed as a d-wave order parameter admixed with an extended s-wave component $s'=\cos q_x + \cos q_y$. 

We emphasize that at the level of the bare action order parameter fluctuations are not momentum- and frequency dependent.  Momentum- and frequency dependent terms will be generated by quantum fluctuations, however, which in turn determine the ordering wave vector in the symmetry broken phase.

\section{Nematic susceptibility in RPA}
\label{sec:RPA}

Within RPA the propagator $D_0(\k,\Omega_n)$ of the $\phi$-field, which equals the nematic susceptibility, takes the form
\begin{equation}
D_0^{-1} (\k,\Omega_n) = s - 2 \lambda^2 \, \Pi_0(\k,\Omega_n) \ ,
\end{equation}
where the factor $2$ arises from the electron spin. The wave vector at which the static susceptibility $D_0(\k, \Omega_n \to 0^+)$ is maximal determines the modulation wave vector of the nematic order parameter in the ordered phase. $\Pi_0(\k,\Omega_n)$ is the bare d-wave polarization function defined by
\begin{equation}
\Pi_0(\k,\Omega_n) = -\sum_{q} G_0(\k+\q,\Omega_n + \omega_n) G_0(\q,\omega_n) \, V^2_{\k,\q+\frac{\k}{2}} 
\label{Pi}
\end{equation}
where $G_0(\k,\omega_n) = (i \omega_n - \xi_\k )^{-1}$ denotes the bare electron Green's function. 

\begin{figure}
\begin{center}
\includegraphics[width=\columnwidth]{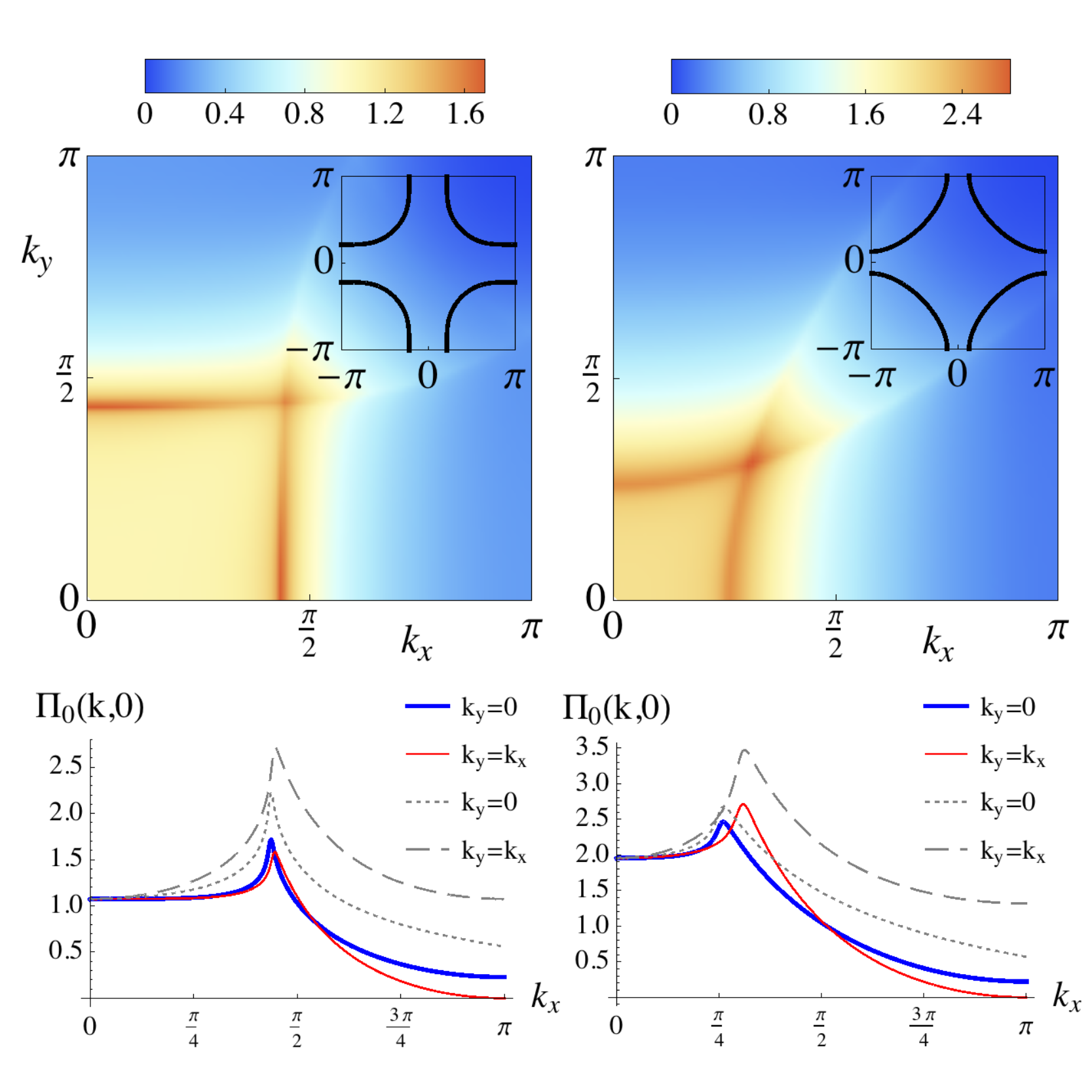}
\caption{(Color online) Static d-wave polarization function $\Pi_0(\k,0)$ for different Fermi surface geometries. Left column: $t'/t=-0.32, \ t''/t = 0.128, \ \mu/t=-0.85$ ($8\%$ hole doping relative to half filling), right column: $t'/t=-0.2, \ t''/t = 0.05, \ \mu=-0.75$ ($12\%$ hole doping). The upper graphs show density plots of $\Pi_0(\k,0)$ in one quadrant of the Brillouin zone and the insets indicate the shape of the Fermi surface. The lower graphs show cuts of $\Pi_0(\k,0)$ along the x-axis (blue thick line) and along the Brillouin zone diagonal (thin red line). For sufficiently strong antinodal nesting the maxima of the polarization function are at the axial wave vectors of the form $\k=(\pm Q_0, 0)$ and $(0,\pm Q_0)$, otherwise the maxima are at diagonal wave vectors $\k=(Q_0,Q_0)$ and symmetry related points. The grey dashed and dotted lines in the lower graphs show the d-wave polarization function as computed by Holder and Metzner [\onlinecite{Holder}] for comparison, i.e.~using $V_{\k,\q} = 2 d_\q$ instead of Eq.~\eqref{eq:formfactor}.}
\label{fig:geom}
\end{center}
\end{figure}

Holder and Metzner [\onlinecite{Holder}] computed a similar static d-wave polarization function for electrons on a square lattice using $V_{\k,\q} \sim d_\q$ rather than Eq.~\eqref{eq:formfactor} and showed that the nematic susceptibility is enhanced for momenta $\k$ satisfying the condition $\xi_{(\k+\mathbf{G})/2}=0$, with $\mathbf{G}$ a reciprocal lattice vector. These special momenta connect points on the Fermi surface with parallel tangents and are simple lattice generalizations of $2 k_F$-momenta in systems with a circular Fermi surface.\cite{Altshuler} 
At electron densities higher than van-Hove filling the maximum of the nematic susceptibility is at the diagonal wave vectors $\Q=(\pm Q^\text{AF}_0, \pm Q^\text{AF}_0)$, where the two $2k_F$ lines with $\mathbf{G}=(2 \pi,0)$ and $(0,2 \pi)$ intersect. A similar situation is shown in the right panel of Fig.~\ref{fig:geom}, where we plot the static d-wave polarization function as defined above using Eq.~\eqref{eq:formfactor} as a function of momenta at a temperature $T/t=0.01$. Interestingly, these diagonal wave vectors connect antiferromagnetic hotspots on the Fermi surface. Indeed, after shifting momenta by $(0,2 \pi)$ the condition of the two intersecting $2 k_F$ lines $\xi_{\Q/2} = \xi_{\Q/2+\boldsymbol{\pi}}=0$ is equivalent to the condition which determines the position of the antiferromagnetic hotspots $\xi_{\q_\text{AF}} = \xi_{\q_\text{AF}+\boldsymbol{\pi}}=0$, i.e.~$\Q=2 \q_{AF}$ connects antiferromagnetic hotspots on opposite sides of the Fermi surface with parallel tangents.

Here we want to point out that the momentum along the $2k_F$ lines where the nematic susceptibility is maximal also depends on the curvature of the Fermi surface in the antinodal region around $\k = (0,\pi)$ and symmetry related points in the model with a site-centered Ising-nematic field considered here.  For the case of sufficiently strong antinodal nesting, the maximum is at a wave vector $\Q_0 = (Q_0,0)$ and symmetry related points, which connects opposite antinodal points of the Fermi surface, as indicated in Fig.~\ref{fig:FS}. We show an example of this situation in the left panel of Fig.~\ref{fig:geom}, where we plot the static d-wave polarization function at $\omega=0$ and $T/t = 0.01$ for a different Fermi surface geometry with less curvature in the antinodal region. The cuts along the principal axis and the diagonal of the Brillouin zone show clearly that the maximum is at $\Q_0 = (Q_0,0)$ rather than $\Q=(Q^\text{AF}_0, Q^\text{AF}_0)$. 
We emphasize that the form factor \eqref{eq:formfactor} which depends explicitly on the fluctuation wave vector $\k$  is crucial in order to observe this dependence on the Fermi surface geometry. For $V_{\k,\q} \sim d_\q$ the maximum remains at the diagonal wave vectors in RPA, independent of geometric details of the Fermi surface. 

\begin{figure}
\begin{center}
\includegraphics[width=0.95\columnwidth]{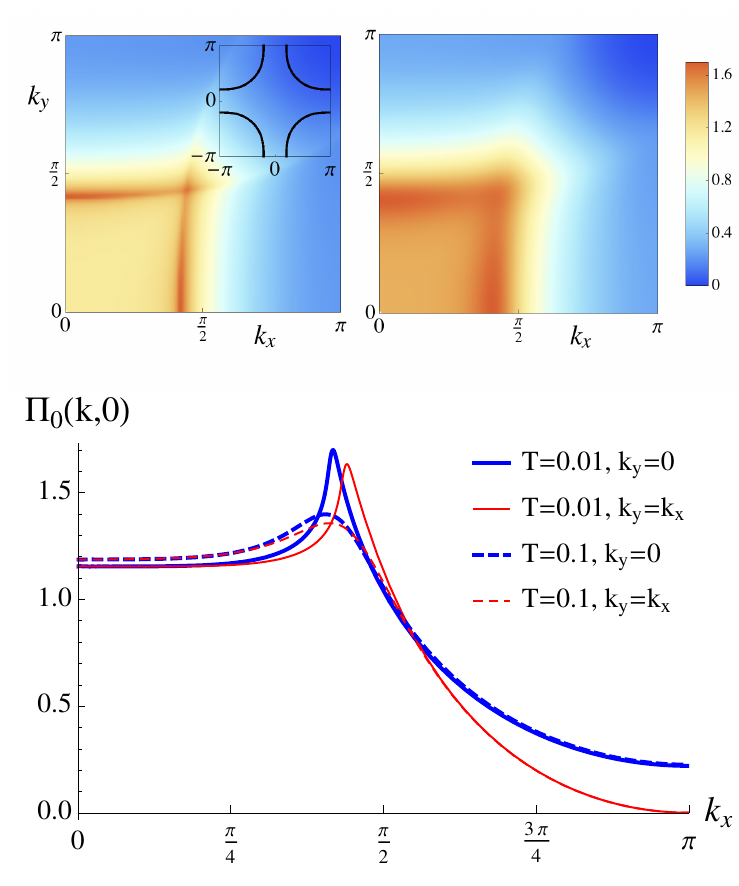}
\caption{(Color online) Density plots of the static d-wave polarization function $\Pi_0(\k,0)$ at temperature $T/t=0.01$ (upper left) and $T/t=0.1$ (upper right). The lower graph shows cuts along the x-axis and the Brillouin zone diagonal. Parameters: $t'/t=-0.3, \ t''/t = 0.1, \ \mu/t=-0.8$ corresponding to $6.6\%$ hole doping.}
\label{fig:RPA_T}
\end{center}
\end{figure}

At finite temperature the $2 k_F$ singularities in the nematic susceptibility are broadened and the maximum is at a slightly smaller wave vector, as shown in Fig.~\ref{fig:RPA_T}, where we compare the static d-wave polarization function at $T/t=0.01$ and $T/t=0.1$. We note that this is no longer the case if effects beyond RPA are taken into account. Within Eliashberg theory the maximum shifts to $\Q=0$ at $T/t=0.1$, as will be shown in Sec. \ref{sec:eliashberg}.

\begin{figure}
\begin{center}
\includegraphics[width=0.9 \columnwidth]{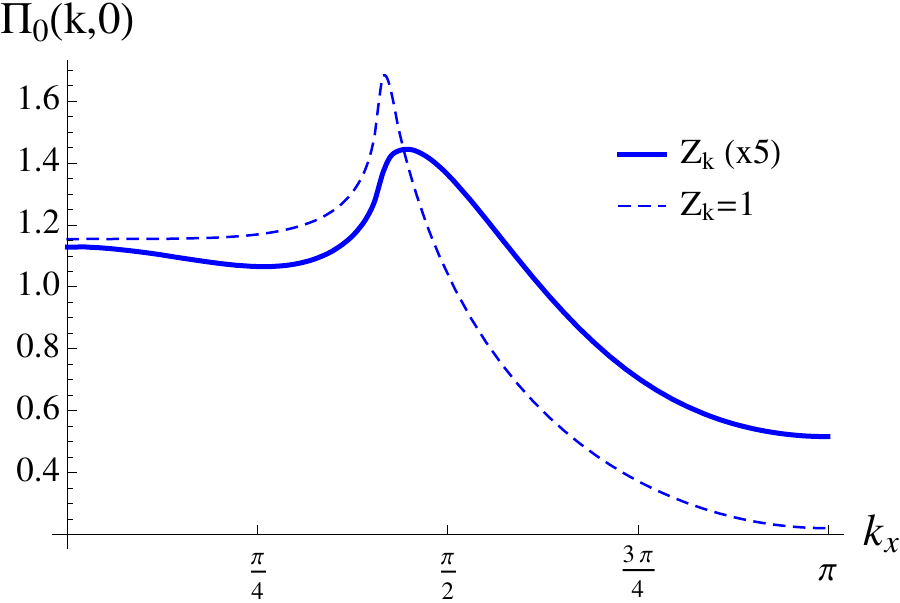}
\caption{(Color online) Static d-wave polarization function $\Pi_0(\k,0)$ at temperature $T/t=0.01$ along the $k_x$ axis ($k_y=0$). The thick solid line shows the result using the renormalized electron propagators \eqref{eq:Gzk} with the quasiparticle residue $Z_\k$ from Eq.~\eqref{eq:Zk} and has been multiplied by a factor of $5$ for better visibility. The thin dashed line shows the result with $Z_\k=1$ for comparison. The reduced quasiparticle residue in the antinodal region shifts the local maximum of the susceptibility from $\Q_0=(1.31, 0)$ to a larger wave-vector $\Q=(1.40, 0)$, indicated by the red arrow in Fig.~\ref{fig:FS}. Parameters: $t'/t=-0.3, \ t''/t = 0.1, \ \mu=-0.8$.}
\label{fig:Zk}
\end{center}
\end{figure}

As mentioned in the introduction, the experimentally observed CDW ordering wave vectors do not seem to connect the Fermi surface at the antinodal points. Rather, it has been suggested that the ordering wave vectors connect the tip of the Fermi arcs in the pseduogap phase.\cite{Comin2014} 
Even though the RPA results above give the correct qualitative form of the ordering wave vectors if the Fermi surface is sufficiently nested in the antinodes, there is clearly something missing.
Indeed, nematic fluctuations give rise to a strong interaction between electrons in the antinodes, which in turn leads to a suppressed quasiparticle coherence for electrons with momenta in the antinodal region. The nematic susceptibility is very sensitive to quasiparticle coherence, however, and thus it is reasonable to expect that the maximum of the susceptibility shifts to larger wave vectors connecting points on the Fermi surface that are no longer perfectly nested, but have a larger quasiparticle residue. 
For this reason it is important to study the susceptibility beyond RPA, where such effects are not taken into account. The next section contains detailed results for the nematic susceptibility within Eliashberg theory, but for now we keep our discussion at a phenomenological level. The simplest approach to include electronic quasiparticle coherence in RPA is to compute the d-wave polarization function in Eq.~\eqref{Pi} with renormalized electron propagators
\begin{equation}
G(\k,\omega_n) = \frac{Z_\k}{i \omega_n - \xi_\k} \ .
\label{eq:Gzk}
\end{equation}
The quasiparticle residue $Z_\k$ is unity along the nodal lines $k_x = \pm k_y$, where the electrons do not interact with nematic fluctuations, and falls off towards the antinodes. A very simple phenomenological parametrization of the quasiparticle residue consistent with the square lattice symmetry takes the form
\begin{equation}
Z_\k = 1 - (\cos k_x - \cos k_y)^2/4 \ .
\label{eq:Zk}
\end{equation}
In Fig.~\ref{fig:Zk} we show the static d-wave polarization function computed with a quasiparticle residue of this form. As anticipated, the local maximum of the nematic susceptibility indeed shifts to larger wave vectors, indicated by the red dashed arrows in Fig.~\ref{fig:FS}. Including $Z_\k$ the wave vector of the maximum changes from $\Q=(1.31, 0)$ to $\Q=(1.40, 0)$, which connects points on the Fermi surface away from the antinode with tangents that are no longer perfectly parallel. 

It is important to note that the global maximum of the susceptibility changes from axial to diagonal wave vectors $\Q=(Q^\text{AF}_0, Q^\text{AF}_0)$ in this simple scenario considered here. This is not surprising, because the quasiparticle residue at the antiferromagnetic hot-spots is larger in the ansatz \eqref{eq:Zk} and the diagonal wave vector still connects points on the Fermi surface with parallel tangents. Within Eliashberg theory, where the damping of electronic quasiparticles along the Fermi surface is determined selfconsistently, this is not necessarily true, as will be shown in the next section.

\section{Eliashberg theory}
\label{sec:eliashberg}

We now turn to a self-consistent one loop calculation of the d-wave polarization function $\Pi(\k,\Omega_n)$ and the electronic self-energy $\Sigma(\k,\omega_n)$. In the spirit of Eliashberg theory vertex corrections are neglected and we comment on the validity of this approximation later. 
Writing the full electron and nematic fluctuation propagators as
\begin{eqnarray}
G^{-1}(k) &=& i \omega_n - \xi_\k-\Sigma(k) \\
D^{-1}(k) &=& s - 2 \lambda^2 \Pi(k) \ ,
\end{eqnarray}
the coupled Eliashberg equations for the electronic and bosonic self-energies take the form
\begin{eqnarray}
\Sigma(k) &=& \lambda^2 \sum_{q} G(k-q) D(q) \, V^2_{\q,\k-\q/2} \label{eq:sigma} \\
\Pi(k) &=& - \sum_{q} G(k+q) G(q) \, V^2_{\k,\q+\k/2}  \label{eq:pi} \ ,
\end{eqnarray}
where we used the abbreviation $k=(\omega_n,\k)$ again.
We solve these coupled equations numerically after analytic continuation to real frequencies by a fixed point iteration on a grid of up to $210 \times 210$ points in the first Brillouin zone and $201$ points along the frequency axis. For all results shown in this section the coupling constant is chosen as $\lambda/t = \sqrt{0.1} \simeq 0.316$ and the gap $\tilde{s} = s-2 \lambda^2 \max_{\k} \text{Re} \, \Pi (\k,0)$ of the nematic fluctuation propagator $D(\k,\omega)$ is fixed at the value $\tilde{s}/t=0.01$ at a temperature $T/t=0.01$ (we do not fix the gap at $\tilde{s}/t=0$ at $T=0$ for numerical reasons). At higher temperatures the boson gap is determined self-consistently and at $T/t=0.1$ we obtain $\tilde{s}/t = 0.022$. Furthermore, we fixed the renormalized chemical potential $\tilde{\mu} = \mu - \text{Re} \, \Sigma(k_F,0)$ at the nodal point to avoid large shifts of the Fermi surface.

\begin{figure}
\begin{center}
\includegraphics[width=0.95\columnwidth]{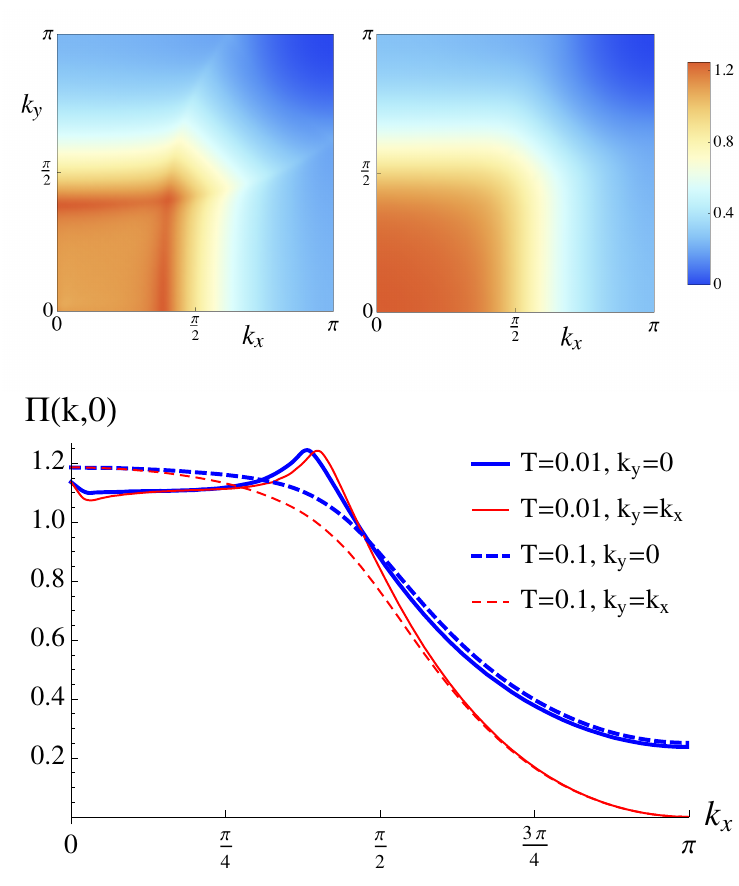}
\caption{(Color online) Density plots of the static d-wave polarization function $\Pi(\k,0)$ computed within Eliashberg theory at a temperature $T/t=0.01$ (upper left) and $T/t=0.1$ (upper right). The lower panel shows cuts through $\Pi(\k,0)$ along the x-axis and the Brillouin zone diagonal. Parameters: $t'/t=-0.3, \ t''/t = 0.1, \ \tilde{\mu}/t=-0.85$.}
\label{fig:eliashberg}
\end{center}
\end{figure}

\subsection{Nematic susceptibility}

In Fig.~\ref{fig:eliashberg} we plot the static d-wave polarization function $\Pi(\k,\omega_n \to 0^+)$ for two different temperatures as a function of momenta. 
In contrast to the RPA results a substantial broadening of the $2k_F$ singularities is already visible at low temperatures $T/t=0.01$, which we attribute to the the reduced quasiparticle coherence of electrons close to the antinodes, as discussed earlier. The maximum of the static nematic susceptibility remains at the axial wave vectors $\Q=(1.20, 0)$ and symmetry related points. We note, however, that it is almost degenerate with the local maxima at the diagonal wave vectors, where the susceptibility is a quarter percent smaller. 

The wave vector $\Q=(1.20, 0)$ is slightly larger than the antinodal nesting wave vector and connects points on the Fermi surface away from the antinodes. Indeed, we determined the antinodal nesting wave vector from the maximum of the electron spectral function $A(\k,\omega) = - \frac{1}{\pi} \text{Im} \, G(\k,i \omega_n \to \omega + i 0^+)$ at $T/t=0.01$ and zero frequency to be $\Q_0=(1.18, 0)$ and the wave vector $\Q$ thus connects hotspots at $\k=(\pm0.60, 2.79)$ on the Fermi surface. 

Interestingly, the $2k_F$ singularities are no longer visible at higher temperatures above $T/t \simeq 0.1$ and the nematic susceptibility is maximal at $\k=0$ instead. The appearance of such a preemptive $\k=0$ nematic order has been discussed earlier in the context cuprates and iron-pnictides.\cite{Wang,Schmalian}

 \begin{figure}
\begin{center}
\includegraphics[width=\columnwidth]{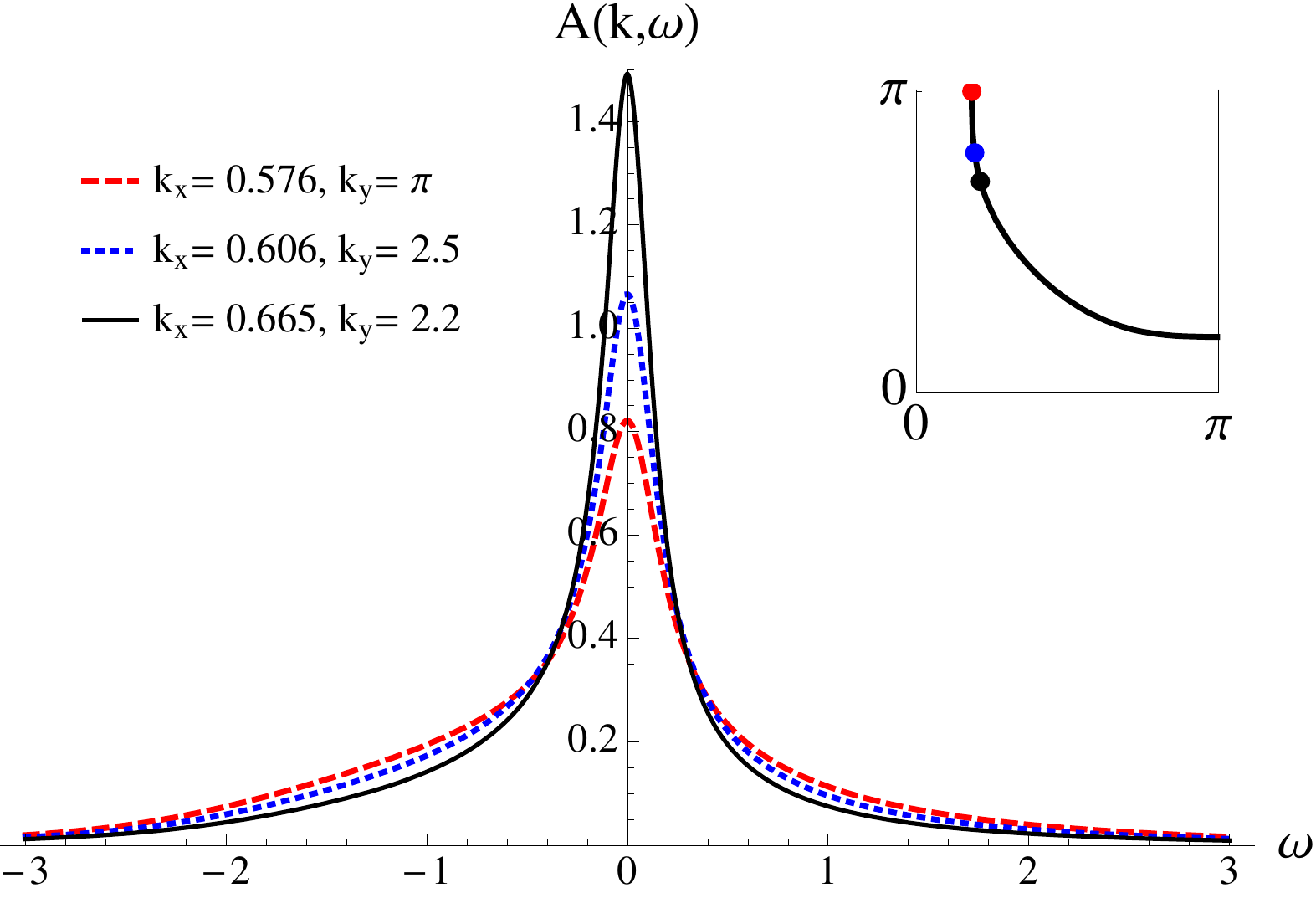}
\caption{(Color online) Electron spectral function $A(\k,\omega)$ at a temperature $T/t=0.1$, plotted as a function of frequency for three different momenta along the Fermi surface. The inset indicates the position of the three momenta on the Fermi surface. Parameters are the same as in Fig.~\ref{fig:eliashberg}.}
\label{fig:specfunc}
\end{center}
\end{figure}

\subsection{Electron spectral function}

Finally we present results for the electron spectral function $A(\k,\omega)$ within Eliashberg theory. In Fig.~\ref{fig:specfunc} we plot $A(\k,\omega)$ at a temperature $T/t=0.1$ as function of frequency for three different momenta along the Fermi surface. The spectral function exhibits a peak at the Fermi energy and its width increases towards the antinodes. There is clearly no pseudogap, i.e.~no suppression of spectral weight at the Fermi energy close to the antinodes. Similar results were obtained already in Ref.~[\onlinecite{Yamase2012}]. It has been argued, however, that the interplay between CDW and d-wave superconducting fluctuations leads to the formation of a low-temperature pseudogap.\cite{Allais2014}

\subsection{Validity of Eliashberg theory}

\begin{figure}
\begin{center}
\includegraphics[width=0.5 \columnwidth]{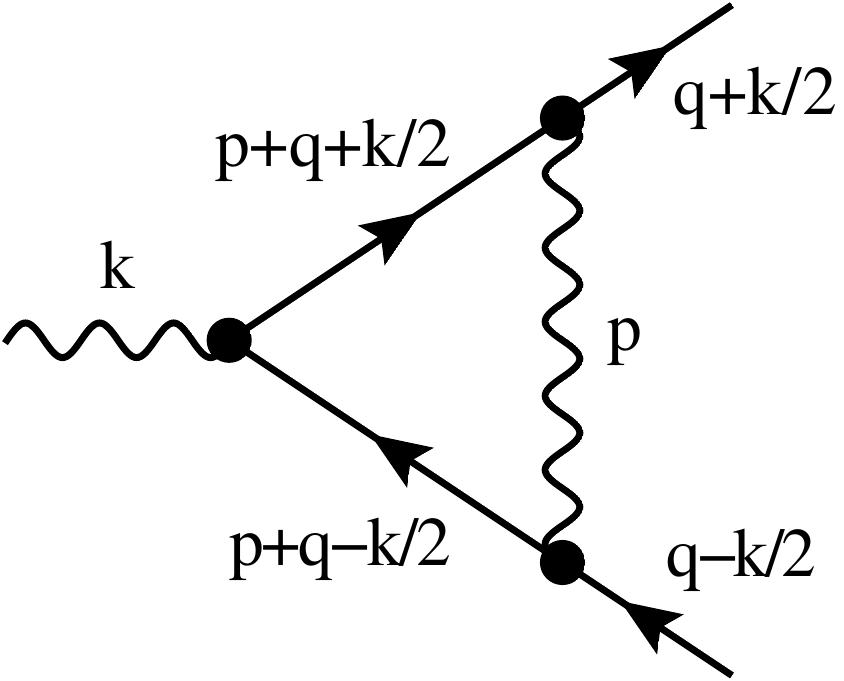}
\caption{Lowest order vertex correction which is neglected in Eliashberg theory. Solid lines denote the electron propagator $G(k)$, wiggly lines represent the nematic fluctuation propagator $D(k)$.}
\label{fig:vertex}
\end{center}
\end{figure}

The validity of the Eliashberg approximation for two-dimensional metals close to a $\k=0$ nematic critical point has been analysed in detail in Ref.~[\onlinecite{Rech}]. One of the requirements there is that the effective four-fermion interaction $\lambda^2 D(\Q,0)$ is small compared to the bandwidth, i.e.~$\lambda^2/\tilde{s} \ll t$ in order for the leading order vertex correction shown in Fig.~\ref{fig:vertex} to be small. Here we want to emphasize one important difference for the case of nematic fluctuations at the axial wave-vector $\Q=(Q_0, 0)$. In fact, the scattering geometry in this case is such that one of the internal electron propagators is always off-shell and the first order vertex correction is small even in a regime where $\lambda^2/\tilde{s} > t$ as considered here. Indeed, setting $\p=\k=\Q$ in the diagram in Fig.~\ref{fig:vertex} one can easily see that the electron with momentum $\q+3\Q/2$ is far away from the Fermi surface if the incoming and outgoing electrons are at the hotspots connected by $\Q$.

To be more explicit, the vertex correction shown in Fig.~\ref{fig:vertex} takes the form
\begin{eqnarray}
\Gamma^{(1)}_{\k,\q} &=& \lambda^3 \sum_p G(p+q+k/2) G(p+q-k/2) D(p)  \notag \\
&& \ \times V_{\k,\q+\p} V_{\p,\q+\frac{\p-\k}{2}} V_{\p,\q+\frac{\p+\k}{2}}  \ .
\end{eqnarray}
Note that $G(k)$ and $D(k)$ denote the fully dressed electron and fluctuation propagators obtained within Eliashberg theory. In order to compute the renormalization of the coupling constant $\lambda$ we project the vertex correction $\Gamma^{(1)}_{\k,\q}$ at zero external frequencies onto the d-wave form factor $V_{\k,\q}$ from Eq.~\eqref{eq:formfactor} and obtain
\begin{equation}
\frac{\delta \lambda}{\lambda} = \frac{1}{2 \lambda} \sum_{\k,\q} V_{\k,\q} \, \Gamma^{(1)}_{\k,\q} \ ,
\end{equation}
where the factor $1/2$ is for normalization. Computing the integral we obtain $\delta \lambda/\lambda = -0.255$ at $T/t = 0.01$ and $\delta \lambda/\lambda = -0.318$ at $T/t = 0.1$.
Note that these vertex corrections are relatively small even though the four-fermion interaction $\lambda^2/\tilde{s}$ is larger than the bandwidth $t$ by a factor of $10$ for the parameters used here.

\section{Conclusions}

We presented numerical results for the nematic susceptibility of electrons on a square lattice below half filling. In contrast to models of electrons coupled to a bond-centered Ising-nematic order parameter field, the model used here exhibits a site-centered field, which can be interpreted as a deformation of a Zhang-Rice singlet centered on the copper sites in the context of hole-doped cuprates. Within RPA the nematic susceptibility is maximal either at diagonal-wave vectors connecting antiferromagnetic hotspots on the Fermi surface, or at the antinodal nesting wave vectors if the curvature of the Fermi surface at the antinodes is sufficiently small. In order to observe this dependence on the Fermi surface geometry it is crucial that the Ising-nematic order parameter has a form factor with $d_{x^2-y^2}$ symmetry which depends explicitly on the fluctuation wave vector. We argued that the strong coupling between electrons and nematic fluctuations close to the antinodes reduces the electronic quasiparticle coherence and shifts the maximum of the nematic susceptibility to larger wave vectors at low temperatures. Results from an Eliashberg-type computation support this picture but indicate that this shift is rather small and can hardly account for the experimentally observed CDW ordering wave vectors in materials such as Bi$2201$, where the CDW ordering wave vector is roughly twice as large as the antinodal nesting wave vector.\cite{Comin} Including the effect of d-wave superconducting fluctuations potentially increases this shift of the ordering wave vector, because such fluctuations reduce the electronic spectral weight at the Fermi level in the antinodal region. We leave this problem open for future study. 

Moreover, we showed that the nematic susceptibility is maximal at zero wave vector above temperatures on the order of $T/t=0.1$. This seems to be in agreement with experiments on underdoped cuprates, where observations of intra-unit cell nematic order have been reported below the pseudogap temperature, which is parametrically larger than reported onset-temperatures of CDW order. It is important to emphasize, however, that the gradual onset of CDW order makes it difficult to determine such an onset-temperature and at present it is not clear if a regime with intra-unit cell nematic order exists above the onset of CDW order in the cuprates.\cite{Julien}
 
Lastly, even though the wave vectors obtained in the Eliashberg calculation are in qualitative agreement with experiments on underdoped cuprates, the simple scenario of electrons close to a nematic quantum critical point clearly doesn't describe the pseudogap behavior, as we do not observe a suppression of electronic spectral weight at the Fermi energy in the antinodal region.

\acknowledgements

We thank J.~Bauer, D.~Chowdhury and M.-H. Julien for very helpful discussions and comments on the manuscript, as well as S. Sachdev for discussions on related topics. This work was supported in part by the Austrian Science Fund (FWF) under grant no.~J 3077-N16 and SFB FOQUS, as well as by the ERC-Synergy Grant UQUAM.


\begin{thebibliography}{999}
\bibitem{footnote}
Van-Hove filling is defined as the electron density where the modulus $Q_0$ of the antinodal nesting wave vector shown in Fig.~\ref{fig:FS} vanishes. In this work we are only interested in situations where $Q_0$ is finite, i.e.~the electron density is larger than van-Hove filling.
\end{thebibliography}
\end{document}